\documentclass[prc,aps,eqsecnum,epsf,amssymb]{revtex4}
\usepackage{graphicx}

\newcommand{\bmq}{{\mbox{\boldmath $q$}}}

\begin{document}
\preprint {WIS 04/Nov 22-DPP}
\date{\today}
\title{A simple qualitative description of EMC ratios $\mu^A$ for 
$0.2 \lesssim x\lesssim 1.5$ and some sample calculations.}
\author{A.S. Rinat and M.F. Taragin}
\address{Weizmann Institute of Science, Department of Particle Physics,
Rehovot 76100, Israel}
\author{M. Viviani}
\address{INFN, Sezione Pisa and Phys. Dept., University of Pisa, I-56100,
Italy}

\begin{abstract}

We study EMC ratios on the basis of a relation between Structure Functions 
(SF) for a nucleus and for a  nucleon, which is governed by a SF 
$f^{PN,A}(x,Q^2)$ of an unphysical nucleus, composed of point-nucleons. 
We demonstrate that the characteristic features of EMC ratios $\mu^A$ 
are determined by the above $f^{PN,A}$ and the SF of free
nucleons. We account for the positions of the points $x_{1,2}$ in 
the interval $0.2\lesssim x \lesssim 0.9$, where $\mu^A(x,Q^2)$=1 
and also for the minimum $x_m$ in that interval. We similarly describe the 
oscillations in $\mu^A$ for $Q^2\lesssim (3.5-4.0)\,$GeV$^2$ 
in the Quasi-Elastic peak region $0.95\lesssim x\lesssim 1.05$ 
and for its subsequent continuous increase  up to $x\approx 1.4$. 
Finally we compute $\mu^A$ over the entire range 
above for $A=^4$He, C, Fe and Au and several $Q^2$ values. The results 
are in reasonable  agreement with both directly measured and indirectly 
extracted data.

\end{abstract}

\maketitle

\hangindent=1in
\hangafter=-4
{\it The literature on  the EMC effect reads like some late 19th
century trilogy. Although spun out over hundreds of pages,
some vague plot just does not reach a  relieving resolution.}
\par

\section{Introduction.}

After the first measurements some 20 years ago, a keen interest developed 
in understanding  EMC ratios $\,\,\mu^A(x,Q^2)=F_2^A(x,Q^2)/F_2^D(x,Q^2)$ 
of Structure Functions (SF) per nucleon of any target $A$ and of the D 
($x$ is the Bjorken variable $0\leq x=Q^2/2M\nu \leq A$; $\,\,\,\nu,Q^2$
are the energy loss and minus the squared 4-momentum transfer; $M$ is the 
nucleon mass). Over the years, re-analysis of older data and a generation of
new experiments have been performed, the latter frequently
at much larger $Q^2$ and often down to very small $x$. Disregarding the 
range $x\lesssim 0.20$, in all EMC experiments with in parallel measured 
$F_2^A, F_2^D$, the Bjorken variable $x$ lies in the $'$classical$'$ EMC 
regime $0.2 \lesssim x \lesssim 0.9$. The canon of direct experiments was 
closed some 10 years ago. Refs. \cite{arn1,gees} report on the status up 
to about a decade ago.

Indirect information on $\mu^A(x\gtrsim 0.9)$ comes from separately measured
inclusive cross sections on targets $A$ and on D. Before 1999 all extractions 
of their ratios for $x\gtrsim 0.9$ have been for relatively low $Q^2$. 
\cite{fs,liuti,omar}. Refs. \cite{day,rock}, \cite{ne3} and \cite{bosted}
contain data for, respectively, $^3$He, $A\ge4$ (SLAC NE3 experiment) and 
on Al. D data for more or less the same kinematics are from Refs. 
\cite{rock1,arnold,schuetz}. Below we shall exploit the more recent JLab 
E89-008 experiment for the extraction of EMC ratios at substantially 
larger $Q^2$ from data on several targets $A$ \cite{arre89,arrd} and on D 
\cite{nicu,arrd}. Finally we mention the Drell-Yan process as an additional
source of information (see for instance Ref.\cite{dy}). 

The combined pools of information lead to the following observations:

i) In the classical regime $0.2 \lesssim x \lesssim 0.9\,\,, |1-\mu^A(x,Q^2)|$ 
hovers between $\approx 0$ and 0.15-0.20 with little $A$ or $Q^2$ dependence. 

ii) In the adjacent range $0.95 \lesssim x \lesssim 1.05$ around the 
quasi-elastic (QE) point
$x=1$, extracted EMC ratios for $Q^2\lesssim 3.5\,$ GeV$^2$ show 
a sharp rise, followed  by an abrupt decrease toward minima around 
$x\approx 1$, the depth of which depends on $A$ and $Q^2$. 

iii) In the $'$deep$'$ quasi-elastic (DQE) region 
$1.05 \lesssim x \lesssim 1.4$, immediately beyond the range ii), EMC 
ratios resume the rise mentioned in ii) with a 
slope increasing with $Q^2$. Those ratios reach maxima of the order 4-7 
and level off, eventually. The very small, and increasingly imprecise 
composing $F_2^{A,D}$ cause considerable experimental scatter in 
$\mu^A$ for the largest $x$.

Attempts to understand the above observations \cite{arn1,gees} 
concentrated primarily on the classical range, where the preferred tool of 
analysis has been the Plane Wave Impulse Approximation (PWIA) 
\cite{cps,cl,omar6,oset}. Different versions did not converge onto an 
unanimously accepted understanding. Some authors conclude
that the crucial ingredients in the PWIA, namely Fermi averaging 
and binding corrections, do not account for the data (see, for
instance, Refs. \cite{sm,arr5}), while others reach the opposite
conclusions \cite{aku1,oset,poles}. From that rather 
frustrating situation sprang alternative, and occasionally far-flung
approaches. We mention the use of Bethe-Salpeter equations for nuclear
vertex functions or bound state wave functions \cite{russ}, medium
modifications of nucleons \cite{arr5}, the introduction of, in the EMC
field, exotic chiral solitons  \cite{sm} and more. 

For two reasons it seems timely to-reopen the nearly stalled discussion. 
First, a new inclusive scattering experiment JLab E03-103 is currently 
running on D, $^3$He and $^4$He targets \cite{arre03}. We shall soon have 
sorely missing accurate EMC data on $^3$He and $^4$He, 
covering a wide range of $x$ which cross the QE point $x=1$. Those will be 
of special interest, because for the lightest nuclei, EMC ratios differ 
from the mainstream of heavier targets, among which the $A$-dependence ot 
those ratios is sizably weaker. Moreover, only for the former class of 
nuclei can one perform accurate calculations.  

The second reason is the recurrently expressed wish for a simple,  
qualitative understanding of EMC ratios. Here one is warned against
pitfalls of over-simplification, for instance in attempts to understand 
the steeply rising $\mu^A$, setting in at $x\approx 0.9$. Off-hand one 
expects $F_2^D(x\lesssim 1)$ to become very small. In fact, the SF $F_2^D$ 
for two non-interacting component nucleons vanishes for $x=1$, and one does
not expect binding effects to be significant. The above is correct, but a  
similar, and  even more pronounced effect occurs for SFs of all heavier 
targets. As a result, deep $minima$, and not maxima occur in EMC ratios 
at $x=1$. 

In spite of the above skepsis, we shall attempt below such a description, 
invoking a relation between nuclear and nucleonic SFs, which is mediated by 
a SF $f^{PN,A}$ of a fictitious nucleus, composed of point-nucleons 
\cite{gr1}. That SF is a covariant generalization \cite{gr2} of a similar 
one in the non-relativistic Gersch-Rodriguez-Smith (GRS) theory for 
inclusive scattering \cite{grs}. It implicitly contains the equivalent of 
Fermi-averaging and binding effects, which are the emphasized ingredients 
of the PWIA, and goes beyond it: it enables a relatively simple computation 
of the dominant Final State Interaction (FSI).

The above covariant GRS approach has been successfully applied to an extensive 
body of inclusive scattering data with $Q^2\gtrsim (2.5-3.0)\,$ GeV$^2$ 
\cite{rt3,rt2,rt1,vkr} and to observables, related to nuclear SF $F_k^A$ 
\cite{rtv}. It is thus natural to study  EMC ratios in that approach. 
Actually, a first version of the model has years ago been shown to 
reasonably account for the measured $\mu^{{\rm Fe}}$ in the classical 
region \cite{rt2}. 

This note is organized as follows:

1) We re-state the relation between nuclear and nucleon SF by means of
$f^{PN,A}$ and list distinct properties of the latter.

2) We demonstrate that virtually all features of EMC ratios
$F_2^A/F_2^d$ in the entire range of our interest $0.20 \lesssim x\lesssim 
1.50$ can be qualitatively understood from the $x,Q^2$, and in particular
from the outspoken $A$ dependence of the above SF $f^{PN,A}$. In the 
$'$classical$'$ EMC regime $ 0.20 \lesssim x\lesssim 0.90$ those
characteristics are the positions of the points $x_{1,2}$, where the EMC 
ratios cross the value 1, including the $A$-dependence of $x_2$ and the 
approximate position of an intermediate minima. In addition, we describe 
how for $Q^2\lesssim(3-4)\,$GeV$^2$ a sharp rise in $\mu^A$, setting 
in for $x\approx 0.9$, abruptly turns into a deep minimum close to $x=1$, 
and continues its rise for $x\gtrsim 1.05$ in a  $Q^2$-dependent fashion. 
For increasing $Q^2\gtrsim (4-5)\,$GeV$^2$ the above minima degenerate into 
some minor structure.

3) In support of the above qualitative considerations, we present results for 
actual calculations and compare those with directly measured and extracted 
EMC data.

4) In conclusion we compare our approach with an approximatively equally 
successful, but less transparent, Distorted Wave Impulse Approximation (DWIA) 
description.

\section{Generalities.}

We start with a previously postulated relation between SF $F_k^{N,A}$ for 
nucleons ($N=p,n$) and a nucleus \cite{gr1,gr2}
\begin{eqnarray}                                                 
  F^A_2(x,Q^2)(\equiv F_2^{A,\delta N}(x,Q^2))&=&\int_x^A\, dz f^{PN,A}(z,Q^2)
   F_2^{\langle N\rangle} \bigg (\frac {x}{z},Q^2\bigg )\ ,
  \label{a1a}\\
        &=&\int_{1/A}^{1/x} du B^A(u,Q^2)F_2^{\langle
        N\rangle}(xu,Q^2)\ ,
   \label{a1b}                                                    
\end{eqnarray}
with
\begin{equation}
   B^A(u,Q^2)=f^{PN,A}(1/u,Q^2)/u^2\ ,\label{a2}
\end{equation}
In both forms the integrands separate $x$ and $A$ dependence. 
Above we use a weighted $p,n$ nucleon SF with ${\delta}N$ the neutron
excess. Thus
\begin{equation} 
  F_2^{\langle N\rangle}=\frac{F_2^n+F_2^p}{2}+\frac{{\delta}N}{2A}
  (F_2^n-F_2^p)\ , \label{a3}
\end{equation}
The connection between the nuclear and averaged nucleon SF above is provided 
by $f^{PN,A}$, which is the SF of a fictitious target $A$, composed of 
point-nucleons. 

The relations~(\ref{a1a}) and~(\ref{a1b}) are exact in the
Bjorken limit and have empirically been shown to hold for finite
$Q^2\ge Q_0^2\approx (2.0-2.5)\,$GeV$^2$ \cite{rt1,rtval}. Also the
PWIA for $F_2^A$ is of the form (\ref{a1a}) with  $f \to f^{PWIA}$.  

Eqs. (\ref{a1a}) and (\ref{a1b}) describe partons which
originate exclusively from  nucleons. 
The same from virtual bosons \cite{lle}, as well as (anti-)screening effects 
\cite{weise}, are negligible for $x\gtrsim 0.2$; we shall restrict
ourselves to that region. Finally, in view of the  relatively high $Q^2$ 
involved, one may neglect the mixture of $F_1^N$ in the integrand in 
Eq. (\ref{a1a}) \cite{atw,sss}.

It will be useful to separate $F^{\langle N\rangle}=F^{\langle N\rangle,NE}+
F^{\langle N\rangle ,NI}$ into components $NE, NI$, which correspond to 
processes in which the nucleon absorbs the exchanged virtual photon 
elastically ($\gamma^*+N\to N$) or inelastically
($\gamma^*+N\to$  hadrons, partons). For our purposes it suffices to recall 
that $F_2^{\langle N\rangle ,NE}=\delta(1-x){\cal G}_2(Q^2)$, with
${\cal G}_2(Q^2)$ some linear combination of squared static nucleon 
form factors. Substitution of the above into Eqs. (\ref{a1a}) trivially 
produces for the corresponding NE part of any nuclear SF
\begin{equation}
  F_2^{A,NE}(x,Q^2)=f^{PN,A}(x,Q^2){\cal G}_2(Q^2)\ .\label{a93}
\end{equation}
The above SF $f^{PN,A}$ are constructed from  many-body target density
matrices, which are diagonal in all except one coordinate. Those can
only be calculated  with precision for the lightest nuclei, $A\le
4$. The computation of $f^{PN,A}$ requires in addition information on
(off-shell) $NN$ scattering (see for instance Ref. \onlinecite{rt2}).

We summarize salient properties of the SF $f^{PN,A}$ \cite{rtv}:

a) The normalized SFs $f^{PN,A}(x,Q^2)$  are smooth functions    
of $x$ and are approximately symmetric around a, $Q^2$-dependent maximum
$x_M^A(Q^2)\approx 1$, close to the QE peak (Fig. 1). We note from 
Eqs. (\ref{a1a}) and~(\ref{a1b}) that the $A$-dependence of nuclear 
SFs is largely governed by the same in $f^{PN,A}$.

b) For given $Q^2$, peak-values of $f^{PN,A}$ strongly decrease with
increasing  
$A$ from D, He to general $A$, and show only few $\%$ differences 
between nuclei with $A>12$. Due to normalization, also their widths  
show marked variations with $A<12$ (Fig. 1).

c) The above peak-values increase with $Q^2$ (see Figs.~2 and~3) and reach
rather slowly an asymptotic limit.

d) Ratios of peak values $f^{PN,A}(x\approx 1,Q^2)/f^{PN,A'}(x\approx 1,Q^2)$
are only weakly $Q^2$-dependent.

The above properties of $f^{PN,A}$ determine those of $B^A$,
Eq. (\ref{a2}), which in Figs. 4 and~5 are displayed for a few targets
and for $Q^2= 3.5, 10\,$GeV$^2$. $B^A$ obviously peaks around
$u\approx 1$. It  
decreases on both sides with increasing $|1-u|$, and due to the factor 
$1/u^2$ in its definition (\ref{a2}), in a more asymmetric fashion
than does $f^{PN,A}$.  In the following, we use $\bar A$ to
specify a generic target with $A\ge 12$.
Inspection of Figs.~4 and~5 shows that the various $B^{\bar A}$
intersect $B^{{\rm D}}$ at $u_i\approx 0.9$ and 1.1, while $B^{{\rm
D}}$ and $B^{^4{\rm He}}$ cross  at values slightly closer to 1. The
above enables $B^A$ to be ordered as function of $A$. Practically
independent of $Q^2$ one  has
\begin{eqnarray}
   B^{\rm D} > B^{{^4}{\rm He}} > B^{\bar A}\ , 
   && 1.1 \gtrsim u \gtrsim 1.0\ ,\label{78a}\\ 
   B^{\rm D} \ll B^{{^4}{\rm He}} \approx B^{\bar A}\ , && 
   u\lesssim 0.9\ ,u\gtrsim1.1\ .\label{78b}
\end{eqnarray}
More details on $B^A$ and other functions to be mentioned are entered in 
Table I.

\section{Characteristic features of EMC ratios.}

Unless stated otherwise, we focus on the usually dominant NI parts of both SFs
in EMC ratios.

\subsection{The classical range $0.2\lesssim x \lesssim 0.90$.} 

I) It is an experimental fact, that the slope of $F_2^{p,D}(x,Q^2)$, which
varies smoothly as function of $Q^2$, vanishes for $x_1\approx 0.18-0.20$ 
(see for instance Ref. \cite{arn}). Since also the $x$-derivative of those
functions around that $x$ is small, standard reasoning justifies in the 
neighborhood of $x_1$ the $'$primitive$'$ determination \cite{prim,cl}
\begin{equation}
  F_2^n(x,Q^2)\approx 2F_2^D(x,Q^2)-F_2^p(x,Q^2)\ . \label{a7}
\end{equation}
Using the above in Eq. (\ref{a3}), one has in that $x$ region
$F_2^D\approx F_2^p \approx F_2^n\approx F_2^{\langle N\rangle}$.
Next, from Eqs. (\ref{a1a}), (\ref{a7}) and property a) in Section II,
one extracts for any $A$ approximate information on the $nucleonic$ components of 
nuclear SFs and their derivatives. For $x' < x_0\approx 0.18 \ll x
\approx 1$ and independent of $A$, one finds
\begin{eqnarray}
   F_2^A(x',Q^2)&=&\int_{x'}^A dz f^{PN,A}(z,Q^2)
       F_2^{\langle N\rangle}\bigg (\frac{x'}{z},Q^2 \bigg )\ ,
      \nonumber\\
      & \approx& F_2^{\langle N\rangle}(x',Q^2) 
         \int_0^A dz f^{PN,A}(z,Q^2)\ ,  \nonumber \\
      & =& F_2^{\langle N \rangle}(x',Q^2)\ ,\label{a8a}\\
    \frac{{\partial}F_2^A(x',Q^2)}{{\partial}x'}&\approx& 
   \frac{{\partial}F_2^{\langle N\rangle}(x',Q^2)}{{\partial}x'}\ ,
   \label{a8b}
\end{eqnarray}
Strictly speaking, Eqs. (\ref{a8a}), (\ref{a8b}) hold for $x'=0$. However, 
the nucleonic parts of $F_2^{\langle N \rangle}(x')$ and 
${\partial}F_2^{\langle N \rangle}(x')/{\partial}x'$ hardly change up to 
$x'\lesssim x_0$, hence for those sufficiently small $x_1$, and
independent of $A$ and $Q^2$, $\mu^A(x_1\approx (0.18-0.20),Q^2)\approx1$. 
This is indeed observed for all EMC data \cite{arn1,gomez}.

II) Next we consider the interval $x_1\lesssim x\lesssim 1$ and focus
on the integrand of Eq. (\ref{a1b}), i.e. the product of the $A$-dependent 
$B^A(u,Q^2)$ and the weighted nucleon SF $F_2^{\langle N\rangle}(xu,Q^2)$.
The former is dominated by the peak region $u\approx 1$, while
$F_2^{\langle N\rangle}$ smoothly decreases with increasing argument $xu$, 
and is practically negligible for $xu\gtrsim 0.80$ (we shall return
below to the physical NE boundary $x=1$).

Whereas the above mentioned ordering in $A$ of $B^A(u)$ depends only 
on $u$, the same for its product with $F_2^{\langle N \rangle}$, and thus of 
the integral $F_2^A$, Eq. (\ref{a1a}), is crucially influenced by  $x$,
for instance in the deep inelastic (DI) $x$-range $0.2 \lesssim x 
\lesssim 0.8$. There the argument $xu$  of $F_2^{\langle N \rangle}$ 
in the $u-$integral~(\ref{a1b}) is usually $\lesssim 0.8$ in regions around 
$u\approx 1\,$, where $B^A$ is large. The above-mentioned $A$-dependence of 
$B^A$ then allows the prediction
\begin{equation}
   \mu^{\bar A}<    \mu^{{^4}{\rm He}} 
    <1\ ,\qquad\qquad  0.2 \lesssim x \lesssim 0.8\ . \label{a9a}
\end{equation}
In order to obtain non-negligible $F_2^{\langle N \rangle}(xu)$ for
increasing $x\gtrsim 0.8$ one needs $u<1\,,$ 
($x>1$ or equivalently $\nu<\nu_{\rm QEP}$). Those $u$ are on the 
elastic, lower $u$ side of the peak, where $B^A$ is appreciably smaller 
than at the peak.
Fig. 6 illustrates the above for $B^A(u)F_2^{\langle N\rangle}(xu)\,$ for 
He, Fe and Au: in each step between the sample values $x=0.15,0.5, 0.85, 1.2$, 
the above  product drops by roughly a factor 10.

Returning to the EMC ratios, we  already established that on the
elastic side $u<1$ of the peak, the functions $B^A$ are ordered
as $B^{\rm D} < B^{{}^{4}{\rm He}} \approx B^{\bar A}$ 
(see also Table I) and therefore expect that 
\begin{equation}
  \mu^{{^4}{\rm He}} \approx 
   \mu^{\bar A} > 1 \ ,\qquad\qquad 0.85\lesssim x \lesssim 0.95\ .
\label{a9b}
\end{equation}
As $x$ grows, the contributing $u$-region tends to contract to
small values of $u$, where $B^D\ll B^A$ and therefore $\mu^A\gg 1$. We
emphasize, that the understanding of the presented orderings do not require 
precise values for the frequently very small values of the involved nuclear
SF, but derive from well-defined, qualitative features of $B^A(u,Q^2)$ 
(i.e. on $f^{PN,A}(x,Q^2)$) and from the simple functional behavior of 
$F_2^{\langle N \rangle}(xu,Q^2)$.

III) From the inequalities (\ref{a9a}), (\ref{a9b}) and the smoothness
of all factors in the integrand, one predicts  a second intersection point 
at $x_2^A\approx 0.85$. In particular, the noted $A$-dependence of $u_i$ 
for which $B^D(u_i)=B^A(u_i)$ causes $x_2^A$ to be larger for $^4$He 
than for $A\gtrsim 12$. The same results from a relativistic PWIA 
calculation \cite{burov}.

We mention here an entirely different approach, where one exploits a sum-rule 
for the nucleonic parts of the involved NI components of  SF (cf. 
Eq. (\ref{a8a}))
\begin{equation}
  \int_0^A \frac {dx}{x}F_2^A-\int_0^2 \frac {dx}{x}F_2^D\approx
  \int_{x_0}^{x_U} \frac {dx}{x}\bigg [F_2^A-F_2^D\bigg ]=0\ .
  \label{a11}
\end{equation}
The approximations in Eq. (\ref{a11}) are based on the widths of $f^{PN,A}$ 
in the SFs, which effectively cut the supports of $x$, and allows the above 
upper and lower limits of integration to be replaced by common $x_U,x_0$. 
As a consequence the difference $F_2^A-F_2^D$ (and in fact $F_2^A-F_2^{A'}$, 
for $any\,\,A,A'$) has to change sign at least once in the interval 
$0.20 \lesssim x \lesssim 0.90$  \cite{rtpdf}, or alternatively their ratio 
has to pass there through 1.

IV) Next we comment on the slopes $s^A(x,Q^2)=\partial\mu^A(x,Q^2)/
{\partial}x$ at $x_{1,2}$. Eqs. (\ref{a8a}) and (\ref{a8b}) imply that 
all SFs and their slopes are about equal in some interval around the small 
$x\approx 0.2$, leading to a small negative, and nearly $A$-independent slope 
of EMC ratios around $x_1$. One estimates 
\begin{equation}
  s^A(x\approx x_1)\equiv \mu^A \bigg [\frac {{\partial}({\rm log}F_2^A)}
  {{\partial}x}-\frac {\partial({\rm log}F_2^D)}{{\partial}x}\bigg ]
  \bigg |_{x_1}\approx s(x\approx x_1)\approx -0.3\ .
  \label{a12}
\end{equation}
While $F_2^A$ hardly depends on $A$ for $x\approx x_2,\,$, its slope 
${\partial}F_2^A(x)/{\partial}x\bigg|_{x \approx x_2^A}$ does so strongly.
Consequently, and in contrast to the same around $x \approx x_1\,\,$, 
$s^A(x\approx x_2^A)$
is positive and large, in agreement with observation \cite{arn1,gomez}.

The above arguments locate the position of a minimum for constant slopes
at $x_m\gtrsim (x_1+x_2^A)/2\approx 0.65$, about as observed. A more 
quantitative estimate for the actual value of $\mu^A(x,Q^2)$ at the minimum 
requires details of the $x$-variation of the slopes $s^A(x)$.

\subsection {The immediate QE peak region $|x-1|\lesssim 0.05$.}

V) We already mentioned that for 
increasing $x$, the rapid fall-off of $F_2^{N,NI}(ux,Q^2)$ in
the integral ({\ref{a1b}) requires $u\ll 0.8$, which Figs.~3 and~4 place in 
the elastic tails of $B^A(u)$. In that region $B^A > B^D$ and
consequently $\mu^A \gg 1$.  

Until this point we considered NI components, which virtually always
dominate nuclear SF and thus the EMC ratios. However, around the QE peak
$x\approx 1$ the NE components are relatively large, in particular for the
lightest targets and for  $Q^2\lesssim (2.5-3.0)\,$GeV$^2$. For a discussion
of their role, it is convenient to introduce relative weights $\gamma^A=
F_2^{A,NI}/F_2^{A,NE}$ in the total $F_2^A=F_2^{A,NI}+F_2^{A,NE}$ and 
additional auxiliary EMC ratios $\mu^{A,NI},\mu^{A,NE}$ for pure NI and NE
components of the contributing SF. One then finds 
\begin{eqnarray}
   \mu^A&=\mu^{A,NI} \frac{\bigg [1+(\gamma^A)^{-1}\bigg ]}{\bigg
   [1+(\gamma^D)^{-1}\bigg ]}\ , &  \label{a13a}\\
    &\approx \mu^{A,NI}\ ,  &  Q^2\gtrsim\, 4\, {\rm GeV}^2\ ;\,\,
    x\lesssim 0.95\ ,\,\, x\gtrsim 1.05\ , \label{a13b}\\
    &= \mu^{A,NE} \frac{\bigg[1+\gamma^A\bigg ]}{\bigg[1+\gamma^D \bigg
     ]}\ , & \label{a13c}\\
    &\approx \mu^{A,NE}\ , & Q^2\lesssim (2.5-3.0)\,{\rm GeV}^2\
    ;\, |x-1|\lesssim 0.05, \,\, x\gtrsim 1.25
    \label{a13d}
\end{eqnarray}
Eq. (3.9) states that, when dominant, NI components are hardly 
perturbed by NE. However, for $A\le 4$ and $Q^2\lesssim (2.5-3.0)\,$GeV$^2$ 
one may have a reversed situation with $\gamma^A\ll 1$. 
In that case, Eq. (\ref{a93}) implies
\begin{equation}
  \mu^A(x\approx 1,Q^2)\approx \mu^{A,NE}(x\approx 1,Q^2)=\frac 
  {f^{PN,A}(x\approx 1,Q^2)} {f^{PN,D}(x\approx 1,Q^2)} \ll 1\ ,
  \label{a14}
\end{equation}
roughly symmetric around $x\approx 1$, as is the case for the SF 
$f^{PN,A}$, see the discussion at point a) of Section IIIA. The sharp
increase beyond the value 1 in $\mu^A$ for $x_2\approx 0.85-0.90$, and
for $Q^2$ in the above range, is thus followed by an equally abrupt
decrease into a deep local minimum. The relative $A$-dependence of the
minimum is read-off from Fig. 1 
\begin{equation}
  \mu^{\bar A}(x\approx 1) < 
  \mu^{{\rm He}}(x\approx 1) \lesssim 1\ ,
  \label{a15}
\end{equation}    
For low $Q^2$, the depth of the minimum is $\approx 0.35$  for $A\ge 12$,
and only $\approx 0.50$ for $A=3,4$.

As long as the NI admixtures in $F_k^A$ are small (cf. Eq. (\ref{a13d})), the
position of the minimum at $x\approx 1$ is only  weakly dependent on $Q^2$, 
because the ratio (\ref{a14}) is so. For increasing $Q^2$, NI components 
grow rapidly relative to NE ones. As a result NI components are never
completely negligible and compete, even when NE is maximal \cite{rtv}. This
affects the  specific NE effect above: the narrow minimum at $x\approx 1$ is
contaminated with NI components, and ultimately vanishes in an asymmetric 
way around $x=1$. The resulting $\mu^A$ for sufficiently large $Q^2$ will
show a practically smooth steep increase from $x\approx 0.85$ on (see Figs. 
7,8,9,10). 

\subsection{The DQE region.}

VI)  Further increasing $x>1$, one can no more have simultaneously 
an argument $xu\lesssim 0.8$  of $F_2^{\langle N \rangle}(xu,Q^2)$, 
in Eq. (\ref{a2}), and $u$ in $B^A(u)$ anywhere close to 1. Consequently
both small factors in the integrand combine to yield strongly reduced  
values for nuclear SF $F_2^A$ (cf. Fig. 6). The crucial 
issue is a reliable prediction for the ratios $\mu^A=F_2^A/F_2^D$.

The approximation $\mu^A\approx\mu^{A,NE}$, Eq. (\ref{a13d}) also holds
for $x\gtrsim 1.25$, where NI components decrease faster in $x$, ultimately 
leading to NE$\gg$NI.

Figs. 4 and~5 show that for sufficiently small $u$ all $B^A$ decrease with a 
characteristic $A$-dependence. For fixed $u$, again the effect is most 
prominent for the D. Compared to it, the decrease is less, and about
similar for He and Fe, and even smaller for the other targets considered. 
Comparison of Figs. 4,5 shows a clear $Q^2$-dependence. Consequently EMC 
ratios rapidly increase for  $x\ge 1.05$ to values far in excess of 1
in an $A$ and $Q^2$-dependent fashion.

This concludes a more than heuristic description. There is little doubt that 
the simple, outspoken $x,A$ dependence of the SF $f^{PN,A}$ for a nucleus of 
point-nucleons describes  all characteristic features of EMC ratios 
in the (deep) inelastic classical regime $0.2 \lesssim x \lesssim 0.95$, 
through the QE region $0.95 \lesssim x \lesssim 1.05$, and up in $x$  into the
DEP region $1.05\lesssim x\lesssim 1.4$. Actual 
quantifications of the above observations are obviously required. 

We close this Section with a remark on generalized EMC ratios. Up to this
point we dealt with EMC ratios of $F_2^A$ and $F_2^D$. Similar considerations 
can be forwarded when comparing $F_2^A$ with $F_2^{\langle N\rangle}$, 
the properly weighted SF of a nucleon. The situation is different for 
generalized EMC ratios $\mu^{A,A'}\,$, $A'\ge 12$, which from the above 
are seen to deviate from 1 by no more than a few $\%$ \cite{arn3}. Again the 
He isotopes occupy a special position. For  instance in the classical regime 
$|1-\mu^{A,^4{\rm He}}|<|1-\mu^{A,^3{\rm He}}|<|1-\mu^{A,D}|$. 
Indeed, data on EMC ratios 
$\mu^{A,^3{\rm He}}$ show all features of $\mu^{A,D}$, but in a more 
temperate fashion \cite{egiyan}.

\section{Input, Results.}

In the following we present computed EMC ratios, which should underscore the 
above qualitative considerations. We start with some input elements.

a) SFs $f^{PN,A}$ for nuclei composed of point-nucleons can only be computed 
with great precision for the lightest nuclei \cite{rt1,vkr}. For heavier 
nuclei approximations are unavoidable and we used the one discussed in Ref. 
\cite{rt3}. For all $A$ we calculated  $f^{PN,A}(x,Q^2)$ from 
a related function $\phi^A(y_G,|\bmq|)$ defined in terms of different
kinematic variables, namely $|\bmq|$, the 3-momentum transfer, and $y_G$, 
the Gurvitz scaling variable \cite{gur}, which is typical for the
underlying, non-perturbative method (see Ref. \cite{rt3})    
\begin{eqnarray}
   y_G^A &=&\frac {2y_G}{1+\sqrt{1+2\nu y_G/(M_{A-1}|\bmq)}}\ ,
  \label{a16a}\\ 
   y_G   &=& M\nu/|\bmq|[1-\langle \Delta\rangle /M-x]\ ,
   \label{a16b}\\
   \delta(x) &\approx & -\frac {\delta\langle \Delta\rangle}{M}\ .
   \label{a16c}
\end{eqnarray}
Above, $y_G^A$ for a recoiling spectator nucleus with finite mass $M_{A-1}$
is in Eq. (\ref{a16a}) expressed in terms of the same, Eq. (\ref{a16b}), for 
an infinite mass spectator. Only for the lightest nuclei does one have to 
retain the recoil correction in Eq. (\ref{a16a}) \cite{gr2,vkr}. 

In expression (4.2) appears an average separation energy of a nucleon 
in the target, which for $^4$He, C, Fe, Au we took as $\langle \Delta\rangle=
20.2, 40, 45, 45$ MeV. Eq. (\ref{a16c}) expresses for given $y_G$ the change 
in $x$, due to the same in $\langle \Delta\rangle$. The above is 
important if $f^{PN,A}$ varies rapidly with $x$, i.e. in the neighborhood 
of $x=1$. A consequence of this fact will be shortly encountered.

b) Since for all applications $Q^2\ge 3.5$ GeV$^2$, we employ for
$F_2^p$ a parametrization of resonance-averaged data \cite{arn}, instead
of one for $F_2^p$ itself for $Q^2\lesssim (4.0-4.5)\,$GeV$^2$
\cite{christy}.
                                                       
c) $F_k^n$ is not-directly accessible and we follow a previously 
introduced method for its extraction from inclusive scattering data on 
targets, where neutrons are bound \cite{rt4}. There one parametrizes the ratio 
\begin{equation}
   C(x,Q^2)=F_2^n(x,Q^2)/F_2^p(x,Q^2)={\sum_{k=0}^2} d_k(Q^2)(1-x)^k
\label{a17}
\end{equation}
and determines a minimal set of parameters in Eq. (\ref{a17}) from the known 
values $C=1$ for $x=0$ and $C=0.75$ for the small $x=0.2$, for which
the primitive relation (\ref{a7}) holds. Beyond the lowest
inelastic threshold, $C(x,Q^2)=0$, except at the physical boundary
$x=1$, where $C$  depends solely on static form factors 
\begin{equation}
  C(x=1,Q^2)=\frac{\bigg (G_E^n(Q^2)\bigg )^2+
  \eta \bigg (G_M^n(Q^2)\bigg )^2}
  {\bigg (G_E^p(Q^2)\bigg )^2+\eta \bigg (G_M^p(Q^2)\bigg )^2}\ .
  \label{a18}
\end{equation}  
The form factors above are those advocated in \cite{bba} from cross section 
data (see also Ref. \cite{arr5}).

At this point we remark on extracted EMC ratios. For relatively low $Q^2$ 
those are overwhelmingly from SLAC NE3 data \cite{ne3}, and those have been 
discussed before \cite{bosted,fs,liuti,omar}. We therefore limit ourself to 
EMC ratios extracted from JLab E89-008 data with $Q^2\gtrsim 3.5\,$GeV$^2$ 
\cite{arre89}, and to which the remarks at the end of a) in this Section 
apply: A
proper analysis of those data for fixed scattering angles $\theta$, for which 
$Q^2$ varies with $x$, requires an interpolation to a few given $Q^2$, 
but there are not enough data points to do so reliably. In the end we 
performed computations for $A$=D, $^4$He, C, Fe, Au at fixed 
$Q^2= 3.5, 5.0, 10\,$ GeV$^2$ and $0.2\lesssim x\lesssim 1.5$. Actually
shown data points for $x\gtrsim 1.1$ may well differ by $\lesssim 10\%$ from
those for the above fixed $Q^2$.

Here  we pose the question, which $F_2^D$ should be used in the denominator 
of $\mu^A$.  Off-hand it seems best to use experimental values. Nevertheless 
we advocate the computed $F_2^D$, which has been calculated in the same 
approach for all $F_2^A$. Those carry therefore the same 'systematic' 
imperfections which may partly cancel in ratios. The choice is of minor 
practical relevance, since the computed $F_2^D$ agree well with the 
parametrized resonance-averaged data \cite{arn}.

In Figs. 7,8,9,10 we display data and  computed EMC ratios for 
$^4$He, C, Fe and Au, at $Q^2=3.5, 5.0, 10.0\,$GeV$^2$ over two separated 
ranges $0.2 \lesssim x \lesssim 0.90$ and $x\ge 0.80$ with different 
vertical scales. For the classical range we used the original data from Refs. 
\cite{amad,ash,bodek} (open circles), while revised results and additional 
data \cite{bcdms,nmc} are given as closed circles.  Occasionally we use
averages as compiled in Ref. \cite{gomez}. In all figures, bars
on data points refer only to statistical errors. One finds confirmed 
the near-insensitivity to $Q^2$, except for $Q^2=3.5\,$GeV$^2$: Ratios of 
slightly irregular $F^A_2$ cause some structure in predictions around $x=0.7$.

In the DI region there is reasonable, but not perfect agreement. At the QE 
peak $x=1$, data show noticeable inelastic NI effects
on the NE ratios $\mu^{A,NE}$, Eq. (\ref{a15}). Beyond, for $x\ge 1.0$ we
show extracted data for actual, non-interpolated $Q^2=(3.4-4.2)\,$ GeV$^2$,
marked by empty squares (see also Refs. \cite{fs,liuti,omar}). Those extracted
for  $Q^2=(4.5-5.3)\,$ GeV$^2$ are shown as filled squares. Theoretical curves 
are for fixed $Q^2=3.5\,, {\rm respectively}\,\,5.0\,$GeV$^2$. 

In the DQE region all data show a strong $Q^2$ and a much more tempered
$A$-dependence, which computations follow to within better than
$\approx 15\%$, except for the largest $x$ for Fe. A glance at Fig. 1 shows
that barely visible changes in the tail can easily make up for, or increase
the difference. In detail:

$^4$He: no data; predicted is a maximum with a subsequent decrease.

C: Predictions follow the steep increase of the data, 
but fall short of them by about $15\%$.

Fe: As for C, but the data (with large error bars!!) continue to 
increase for $x\gtrsim 1.25$, whereas calculations for 
$Q^2\le 5\,$GeV$^2$ show a leveling-off.

Au: Good agreement. No data beyond $x\approx 1.25$, where theory
predicts  maxima for all $Q^2$.

Given the sensitivity and small numbers involved, also the agreement in the 
DQE region to $x\approx 1.2$ is satisfactory. In fact all results are much 
in line with the spelled-out, qualitative predictions with regard to 
intersection points, minima in the classical regime, including their 
$A$-dependence, the structure and $Q^2$-dependence of the minima around 
$x=1$ and a subsequent increase beyond. In both 
regimes $\mu^{{\rm ^4He}}$ is clearly different from other $\mu^A$.
For the He isotopes, the presently running Jlab EMC experiment E03-103 will 
provide a test for the  $Q^2$-dependence of $\mu^A$, in particular for the
predicted gradual fading of the minimum at $x=1$ for increasing $Q^2$.

\section {Discussion, comparison and conclusion.}

In the present note we have studied nuclear SFs $F_2^{A},\,F_2^D$ and their
EMC ratios over the kinematic range $0.2 \lesssim x\lesssim 1.5$. 
All along we have been aware that both composing nuclear SFs in those 
ratios decrease two orders of magnitude between $x\approx 0\,\,
(F_2^A\approx 0.37)$ and $x\approx 1$. In that interval the EMC effect,
i.e. the deviations of EMC ratios from 1 are less than $20\%$ and on 
the average $10\%$. The above requires the accuracy of the computed ratios 
to be better than the size of EMC effect.

Further increasing $x$ to about 1.4-1.5, the involved
SF decrease by another two orders of magnitude and it seems extremely 
hard to fulfill even far less stringent requirements than imposed on the
classical region. The above has not deterred attempts to compute EMC ratios. 

The tool of our analysis has been a postulated relation between nuclear 
and nucleon SFs, linked by $f^{PN,A}$, the SF of an unphysical nucleus, 
composed of point-particles. The computation of the latter requires 
many-particle density matrices, diagonal in all nuclear coordinates except 
one, which are constructed from nuclear ground state wave functions. Such 
a calculation is presently only feasible for the D and the He isotopes and 
approximate methods had to be invoked for $A\ge 12$. In addition one needs 
information and models for off-shell $NN$ elastic scattering.

Calling on characteristic properties of $f^{PN,A}$ as function of $x,Q^2$ 
and $A$ and on the $x,Q^2$-dependence of the properly weighted nucleon SF 
$F_2^{\langle N \rangle}$, we first qualitatively accounted for 
$all$ characteristic features of EMC ratios in the classical regime 
$0.2\lesssim x \lesssim 0.95$, the QE peak area  $0.95 \lesssim x \lesssim 
1.05$, and continuing into the DQE region $\,1.05 \lesssim x
\lesssim 1.4$, including the $A$ dependence of $\mu^A$. The same 
considerations carry over to generalized EMC ratios $\mu^{A,{\rm He}}$, in 
particular for $^3$He, and recent data for the latter confirm those
\cite{egiyan}. 

The above qualitative considerations  lead only to ordering of relevant 
quantities, or alternatively, to inequalities. In the end we performed 
actual calculations of EMC ratios to underscore the above qualitative 
considerations. It is  not surprising that, contrary to the reliable,
qualitative considerations, the very small values of the SF involved do 
not guarantee a similar accuracy in directly calculated results. Nevertheless 
we  obtained reasonable agreement with data in the classical regime 
$0.2 \lesssim x\lesssim 0.9$ and beyond, with extracted EMC ratios
for $0.9 \lesssim x\lesssim 1.25$.

The emphasis above has been on an explanation solely based on the properties 
of the SFs $f^{PN,A}$ and $F_2^{\langle N \rangle}$: We are not 
aware of an alternative simple description, which is not a judgment 
against other approaches. We re-assert the obvious: 
In principle, $'$complete$'$ treatments of various approaches to nuclear SF 
produce identical results. However, in practice  theories are worked out to 
some order in a parameter, which is characteristic for the chosen approach. 
An issue emerges when comparing two results in two approaches to 
{\it different} orders in {\it different} parameters.

At this point we recall a proof, explicitly showing that the GRS and DWIA 
approaches produce similar results if 
both are compared to the $same$ order in a common expansion parameter, e.g.
the lowest order in $NN$ re-scattering \cite{rj}. It does therefore not
come as a surprise to find similar agreement for EMC ratios, computed in 
the afore-mentioned two theories.
 
In spite of the above, it is of interest to note different insights in some 
aspects of EMC ratios, for instance around the minima in $\mu^A(x\approx 1)$. 
The DQE region $x\gtrsim 1$ has repeatedly been discussed 
\cite{fs,liuti}, with as most complete treatment (essentially a DWIA), the 
one by Benhar $et\,al$ \cite{omar}. No particular role is allocated there 
to the QE point. In fact the earlier treatment by Frankfurt $et\,al.$
\cite{fs} allocates the above to the  ratio of the longitudinal 
momentum distributions. In order to reach that result in the DWIA, one has 
first to neglect FSI, which indeed are relatively small around the QEP.
However, in order to reach the longitudinal momentum distribution from the
PWIA, one has to concentrate the spread of the  spectral function into one 
peak. The cited steps are not manifestly equivalent to the dominance of the 
NE contributions in the GRS approach and its consequences. One would also 
have to show
that the above ratio of $Q^2$-independent longitudinal momentum distribution
equals the  $Q^2$ dependent ratio of $f^{PN,A}(x\approx 1,Q^2)$.   

Our last remark regards the treatment of the DQE region. For relatively
low $Q^2$, the EMC ratios have mostly been extracted from  $F_2^A$ (SLAC NE3 
experiment \cite{ne3}) against $F_2^D$ and from a single recent experiment 
of the same against $F_2^{^3{\rm He}}$ \cite{egiyan}. Those data show that,
after an initial increase, the  EMC ratios $\mu^A$ ultimately reach
plateaux of approximately constant values. For increasing $Q^2$, 
Eq. (\ref{a13d}) is  no more valid. As a consequence EMC ratios 
continue to raise, possibly reaching a plateau at a much larger $x$.

In a very simplistic description the appearance of the latter at lower $Q^2$
have been interpreted as due to correlations between the struck and 2,3... 
spectator particles \cite{fs}. It is clearly of interest to see whether 
such a chain of approximations applied to the GRS expressions lead to the 
above-mentioned simple result. Those steps will be elaborated in a 
separate note.

In conclusion, we continue to be amazed and do not fully understand, why 
theoretical EMC ratios appear to agree with data out to large $x$, 
where each participating nuclear SF is extremely small. It is hard to 
believe that the relative uncertainties in the participating $F_2^A$ are
practically $A$-independent and cancel in EMC ratios. To our opinion, 
understanding the above remains a challenge.

\section{Acknowledgment.}

ASR is most thankful to John Arrington, who generously supplied much 
information and also critically read the ms.

\begin{figure}[bth]
\includegraphics[bb=-150 440 567 400,angle=-90,scale=.40]{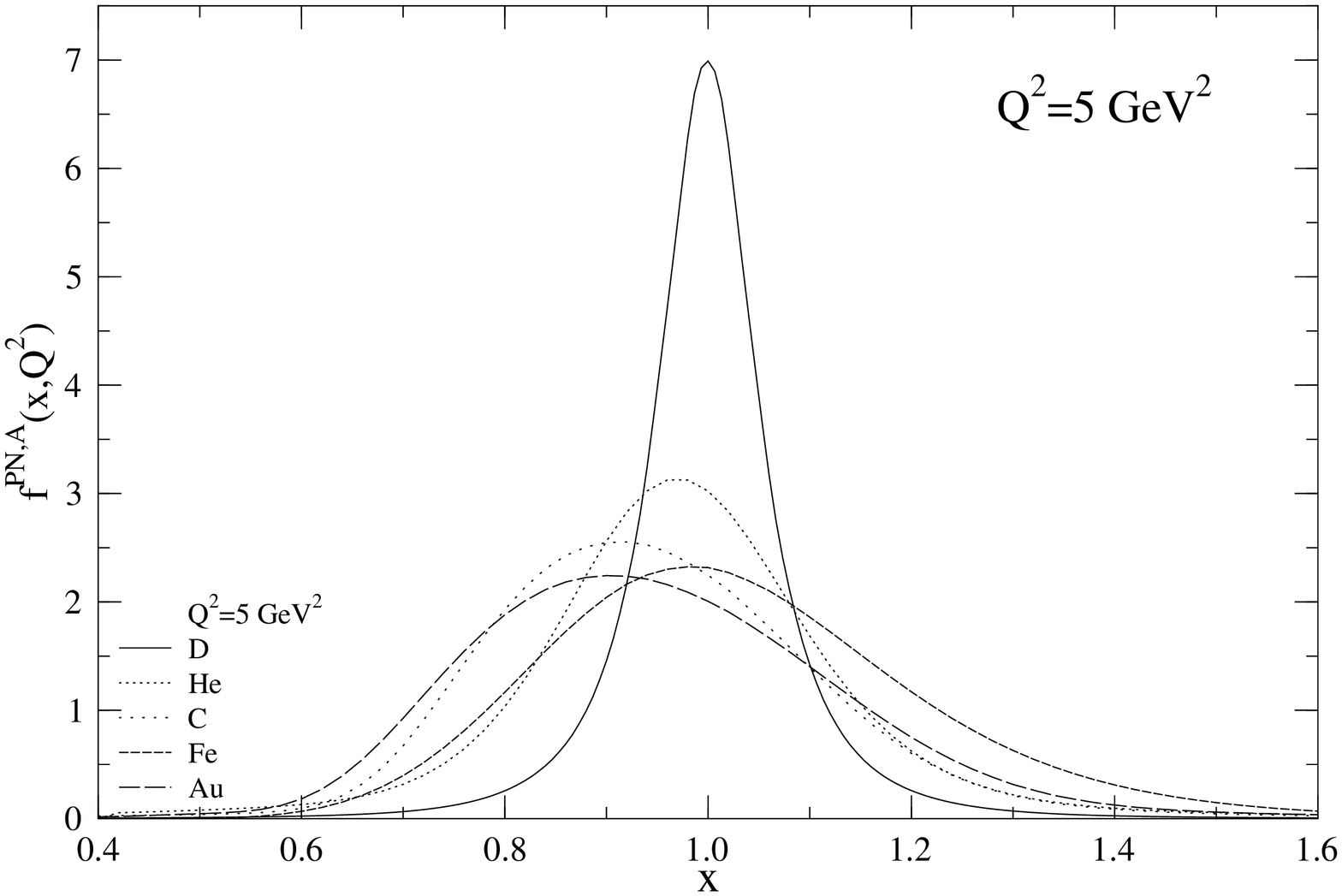}
\caption{ The point-nucleon nuclear SF $f^{PN,A}(x,Q^2)$ for 
D, $^4$He, Fe, C and Au; $Q^2=5\,$GeV$^2$. }
\end{figure}

\begin{figure}[bth]
\includegraphics[bb=-150 440 567 400,angle=-90,scale=.40]{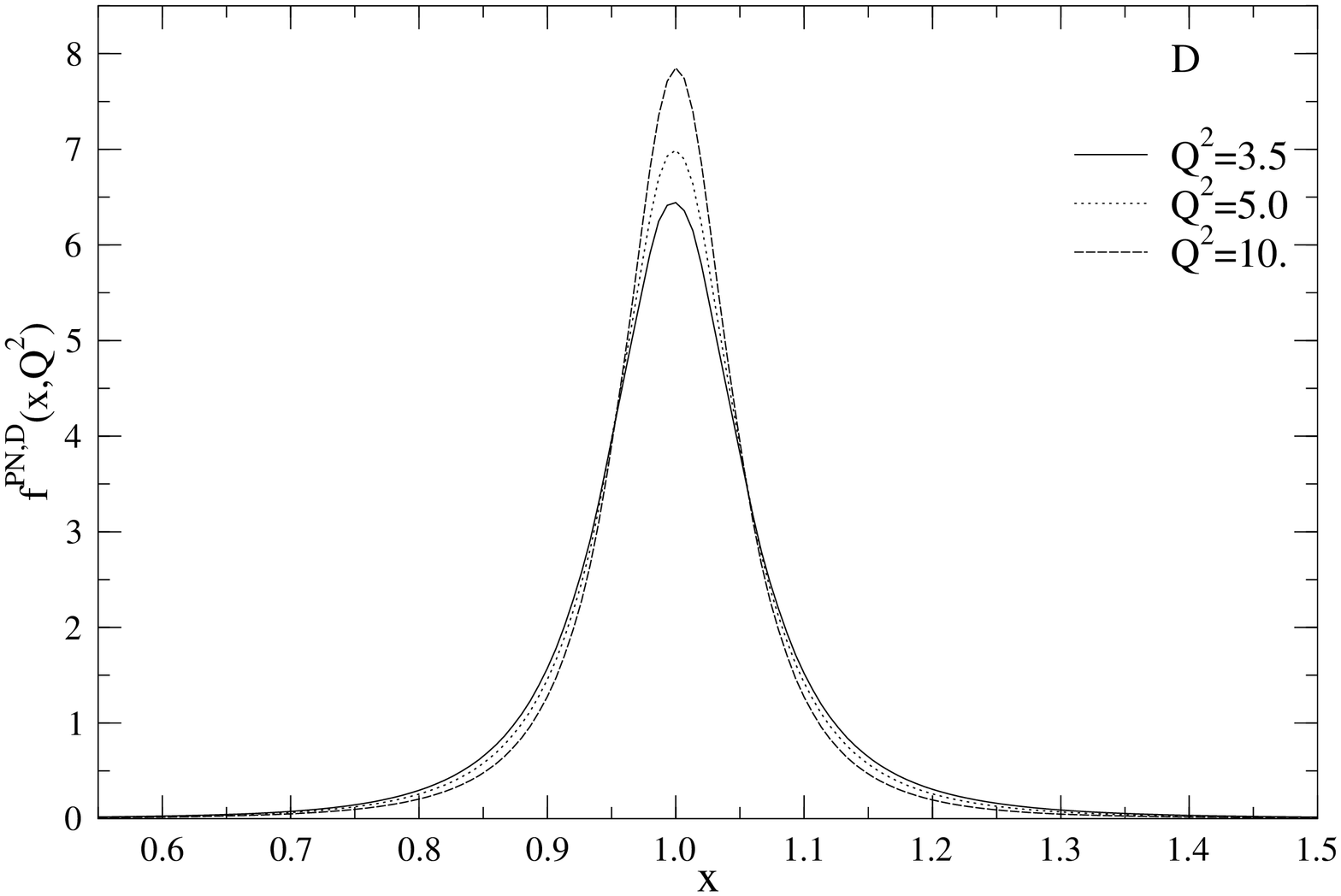}
\caption{Same as Fig. 1 for D; $Q^2=3.5, 5, 10\,$GeV$^2$.}
\end{figure}

\begin{figure}[bth]
\includegraphics[bb=-150 440 567 400,angle=-90,scale=.40]{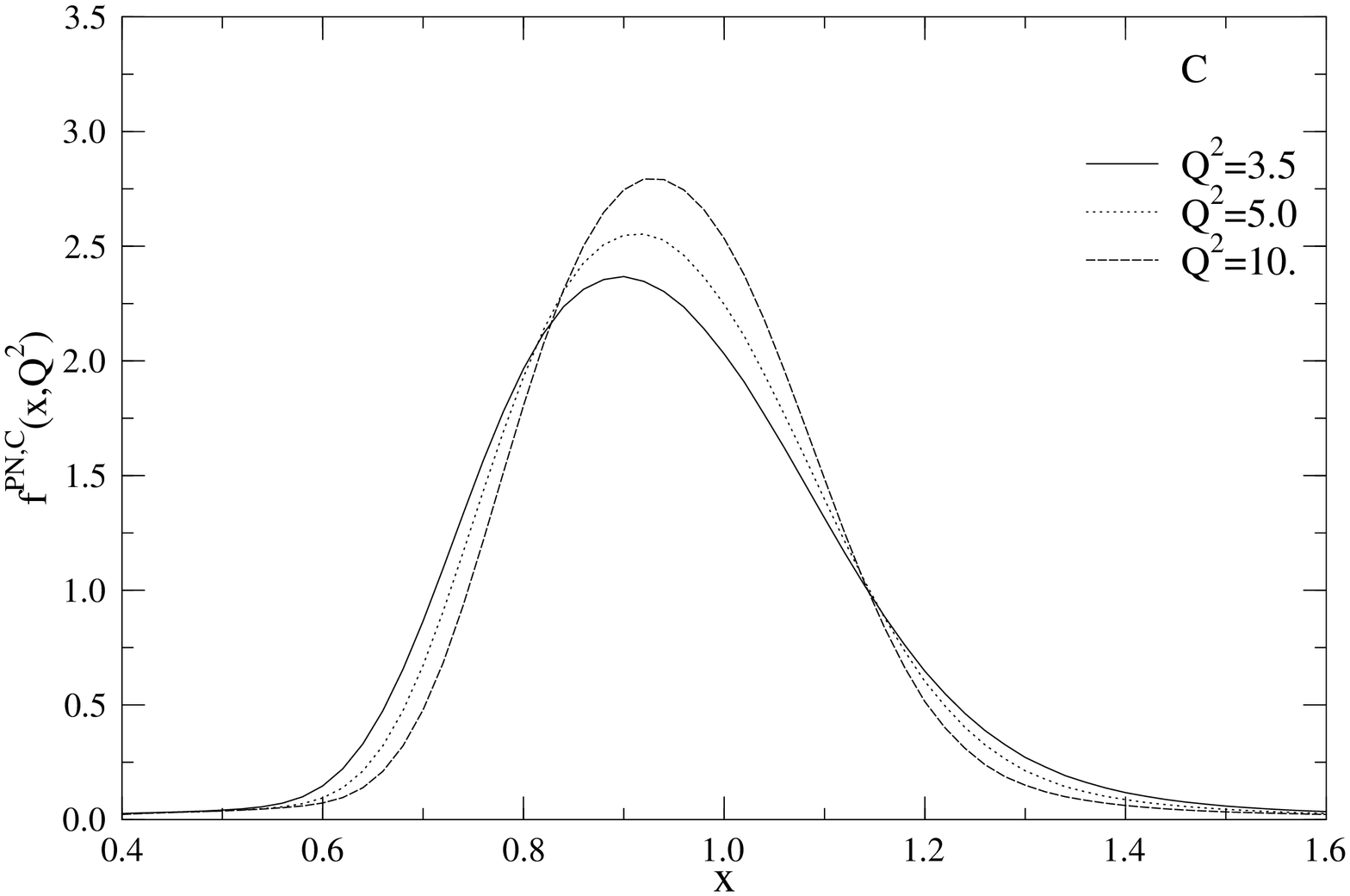}
\caption{Same as Fig. 1 for C; $Q^2=3.5, 5, 10\,$GeV$^2$.}
\end{figure}

\begin{figure}[bth]
\includegraphics[bb=-150 440 567 400,angle=-90,scale=.40]{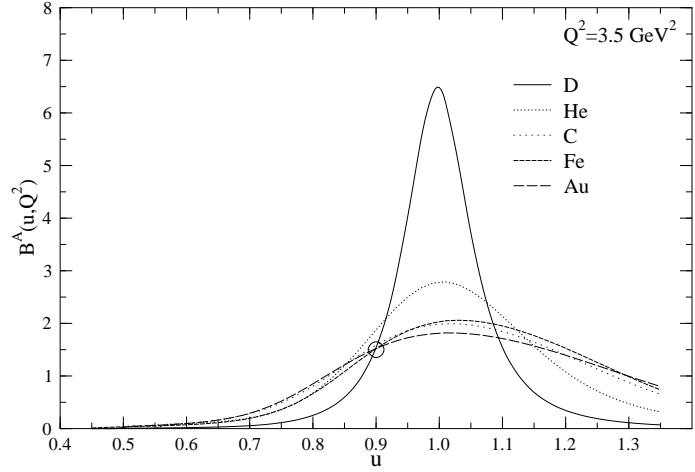}
\caption{The function $B^A(x,Q^2)$, Eq. (\ref{a2}) for our chosen
targets; $Q^2=3.5$GeV$^2$. The circles mark the nearly identical
crossing point $u_i^A$ for D and $A>4$, different from the same for
$^4$He and D.}
\end{figure}

\begin{figure}[bth]
\includegraphics[bb=-150 440 567 400,angle=-90,scale=.40]{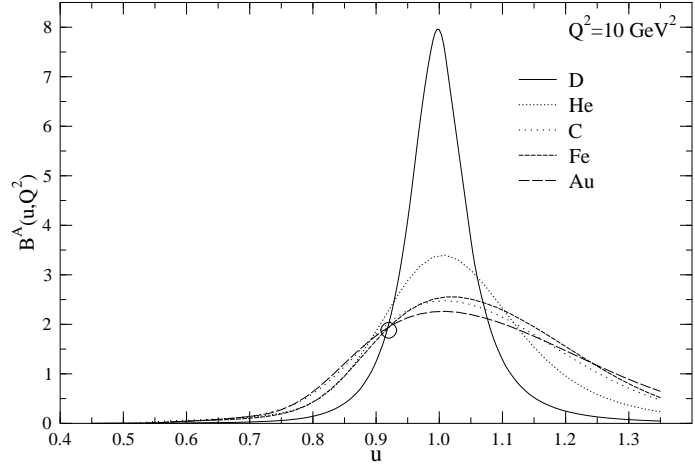}
\caption{Same as Fig.4; $Q^2=10$ GeV${}^2$.}
\end{figure}

\begin{figure}[bth]
\includegraphics[bb=-150 440 567 400,angle=-90,scale=.40]{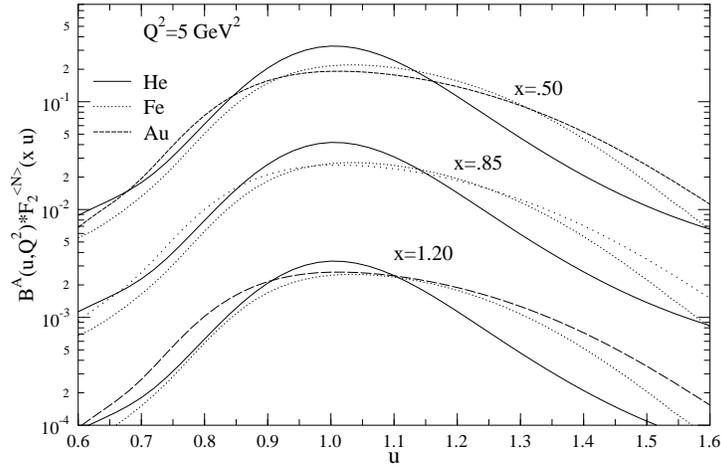}
\caption{The integrand $B^A(x,Q^2)F_2^{\langle N \rangle}(xu,Q^2)$ in Eq. 
(\ref{a1b}), which determines the size of the integrals for $F_2^A$, for 
He, Fe, Au (drawn, dotted and dashed lines) and fixed $Q^2=5.0\,$GeV$^2$. 
The sets of curves show, that the above products decrease about a factor 
of 10 for increasing $x=0.5\to 0.85\to 1.2$.}
\end{figure}

\begin{figure}[bth]
\includegraphics[bb=-150 440 567 400,angle=-90,scale=.40]{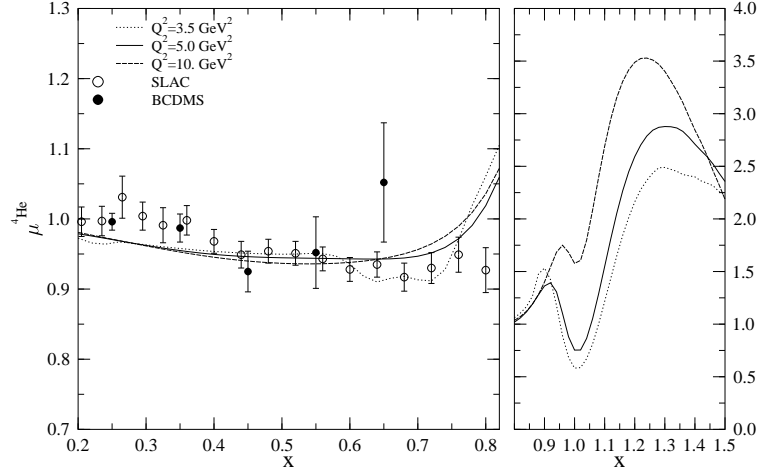}
\caption{$\mu^{^4{\rm He}}$ for $Q^2=3.5, 5.0, 10.0,\,$GeV$^2$. Data in the
classical range are from Ref. \protect\cite{gomez} (open circles) and Ref.
\protect\cite{amad} filled circles. No extracted data exist beyond that range.}
\end{figure}

\begin{figure}[bth]
\includegraphics[bb=-150 440 567 400,angle=-90,scale=.40]{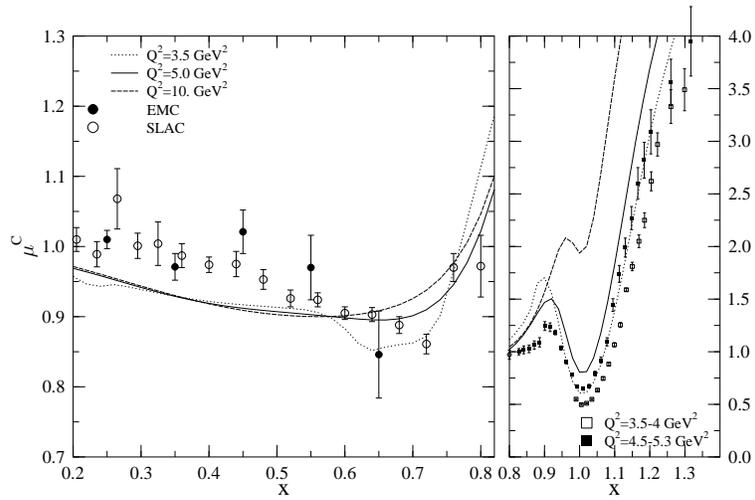}
\caption{Same as Fig. 7 for $\mu^{\rm C}$. Data sets in the classical range 
are from Refs.\protect\cite{ash,gomez}. Extracted data are for varying 
$Q^2\approx 3.4-4.2\,$GeV$^2$ (open diamonds) and for varying $Q^2\approx 
4.5-5.2$ GeV$^2$ (filled diamonds) \protect\cite{arrd,nicu}. Data are too 
sparse for interpolation towards $Q^2=3.5$ and $5.0\,$GeV$^2$ for which 
calculated results are presented.}
\end{figure}

\begin{figure}[bth]
\includegraphics[bb=-150 440 567 400,angle=-90,scale=.40]{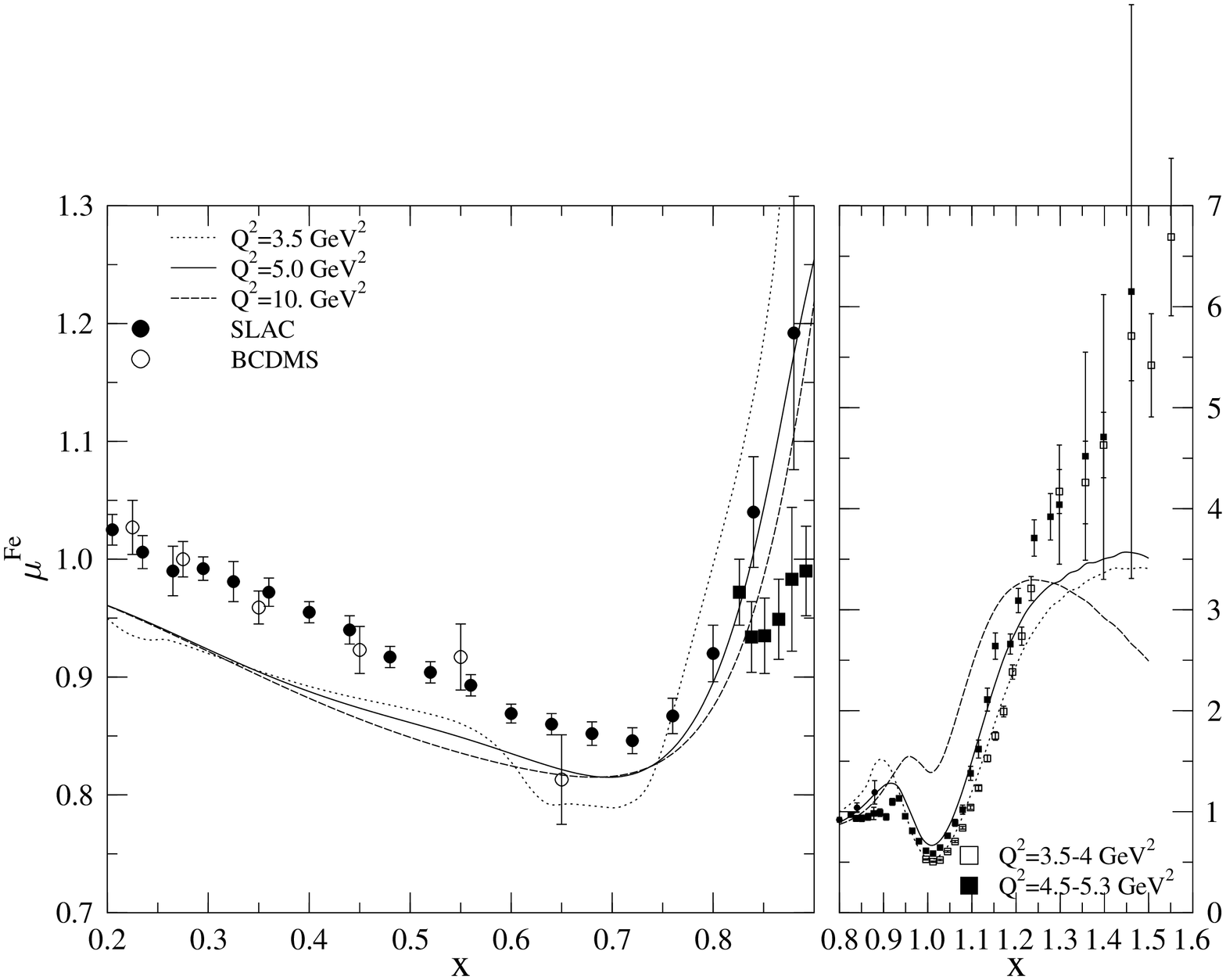}
\caption{Same as Fig. 7 for $\mu^{{\rm Fe}}$ (Data 
from Refs.~\protect\cite{bodek,gomez}; see text in
Section IV).  }
\end{figure}

\begin{figure}[bth]
\includegraphics[bb=-150 440 567 400,angle=-90,scale=.40]{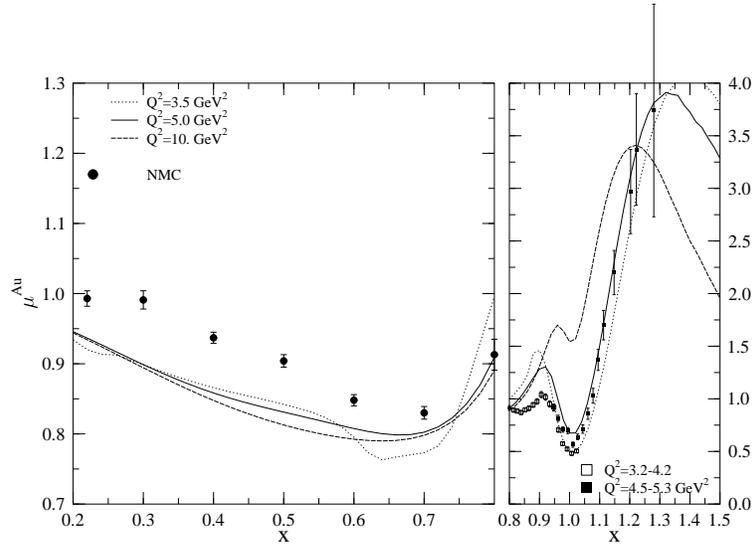}
\caption{Same as Fig. 7 for $\mu^{{\rm Au}}$ (Data 
from Ref.~\protect\cite{gomez}; see text in
Section IV).}
\end{figure}

\begin{figure}[bth]
\includegraphics[bb=-150 440 567 400,angle=-90,scale=.40]{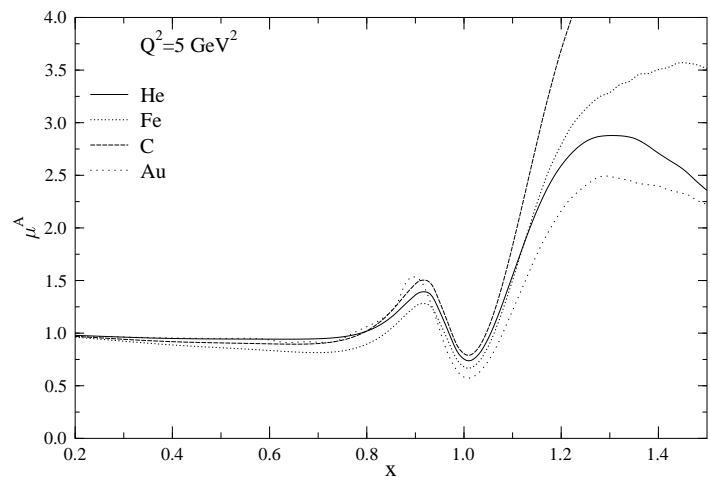}
\caption{A-dependence of computed EMC ratios.}
\end{figure}


\begin{table}[bth]
\caption {Target ordering of $B^A(u,Q^2)$, Eq. (\ref{a93}), as function of
$u,Q^2$. The same for nuclear SF $F_2^A(x,Q^2)$ and EMC ratios
$\mu^A(x,Q^2)$ as function of $x,Q^2$ ($Q^2$ is in units GeV$^2$). When not
distinguishing between different targets, we just use A.}

\begin{tabular}{|c|c|c|c|c|}
\hline
Function &  $u,x$ interval &  $u,x$ interval &  $u,x$ interval
                 &  $u,x$ interval \\
ordered targets &&&&                 \\
\hline
             & $0.6\le u \le 0.9$ &  $0.95\le u \le 1.05$ 
             & $1.05\le u\le 1.1$ & $u\ge 1.1$       \\
\hline
$B^A(u,Q^2\le 10)$ & A $\gg$ D & D $\gg$ He $>$ A
             & He $>$A$\gg$ D  & A$>$ He $\gg$ D                         \\
\hline
             & $0.4\le x \le 0.8$  &$0.85 \le x \le 0.95$ 
             & $0.95 \lesssim x\lesssim 1.05$& $ 1.1 \lesssim x \lesssim 1.5$\\
\hline
$F_2^A(x,3.5\div5.0)$  
                & D $>$ He $>$ C   $\gtrsim$ Fe   &  D $\gtrsim$ A   
                & C $>$ He $\approx$ Fe $\gg$ D  &  C $>$ Fe, He $\gg$ D \\
$F_2^A(x,10.0)$ & D $>$ He $>$ C   $\gtrsim$ Fe  &  C $>$ He $>$ Fe $\gg$D 
                & C $>$ He $>$ Fe  $\gg$ D       &  C $\gg$ He, Fe$\gg$D\\ 
\hline
              & $0.3 \lesssim x\lesssim 0.75$   & $0.85 \le x \le 0.95$
              & $ 0.95\lesssim x \lesssim 1.05$  & $ x\gtrsim 1.05$   \\
\hline
$\mu^A(x,3.5) $ & 1$>$He $>$ A $>$ Fe            & He$\gtrsim$ A$>$1 
                & C  $>$ He$>$A  & C $>$ A $>$Au \\
$\mu^A(x,5.0) $ & 1$>$He $>$ A $>$ Fe            & He$\gtrsim$ A$>$1
                & C  $>$ He$>$A  & C $\gg$ A $>$He  \\ 
$\mu^A(x,10.0)$ & 1$>$He $>$ A $>$ Fe            & He$\gtrsim$ A$>$1 
                & C  $>$ He$>$A  & C $\gg$ A $>$Fe \\
\hline
$Q^2=3.5$ & 1$>\mu^{{\rm He}}>\mu^{{\rm A}}>\mu^{{\rm Fe}}$ 
&   $\mu^{{\rm He}}>\mu^{{\rm A}}>1$ 
&   $\mu^{{\rm C}}>\mu^{{\rm He}}>\mu^{{\rm A}}$
&   $\mu^{{\rm C}}>\mu^{{\rm A}}>\mu^{{\rm Au}}$\\
$Q^2=5.0$ & 1$>\mu^{{\rm He}}>\mu^{{\rm A}}>\mu^{{\rm Fe}}$ 
&   $\mu^{{\rm He}}>\mu^{{\rm A}}>1$
&   $\mu^{{\rm C}}>\mu^{{\rm He}}>\mu^{{\rm A}}$
&   $\mu^{{\rm C}}\gg\mu^{{\rm A}}>\mu^{{\rm He}}$\\
$Q^2=10.0$& 1$>\mu^{{\rm He}}>\mu^{{\rm A}}>\mu^{{\rm Fe}}$ 
&   $\mu^{{\rm He}}>\mu^{{\rm A}}>1$
&   $\mu^{{\rm C}}>\mu^{{\rm He}}>\mu^{{\rm A}}$
&   $\mu^{{\rm C}}\gg\mu^{{\rm A}}>\mu^{{\rm Fe}}$\\
\hline
\end{tabular}

\label{Table I}
\end{table}
\end{document}